\documentclass[aps,prb,twocolumn,superscriptaddress]{revtex4-1}
\usepackage{graphicx,amssymb,amsmath}
\usepackage{natbib}
\usepackage{bm}
\newcommand\BaMnGeO{Ba$_2$MnGe$_2$O$_{7}$}
\newcommand\BaCuGeO{Ba$_2$CuGe$_2$O$_{7}$}

\begin{document}
\title{Instability of Magnons in Two-dimensional Antiferromagnet at High Magnetic Fields}
\author{T. Masuda}
\author{S. Kitaoka}
\author{S. Takamizawa}
\affiliation{International Graduate School of Arts and Sciences,
Yokohama City University, Yokohama, Kanagawa, 236-0027, Japan}

\author{N. Metoki}
\author{K. Kaneko}
\affiliation{Quantum Beam Science Division, JAEA, Tokai, Ibaraki 319-1195, 
Japan}

\author{K. C. Rule}
\author{K. Kiefer}
\affiliation{Helmholtz Znetrum Berlin, GmbH, D-14109 Berlin, Germany }

\author{H. Manaka}
\affiliation{Graduate School of Science and Engineering, Kagoshima University, Kagoshima 890-0065,Japan}

\author{H. Nojiri}
\affiliation{Institute for Materials Research, Tohoku University, Sendai 980-8577, Japan }

\date{\today}

\begin{abstract}
Spin dynamics of the square lattice Heisenberg antiferromagnet, \BaMnGeO , is studied by a combination of bulk measurements, neutron diffraction, and inelastic neutron scattering techniques. 
Easy plane type antiferromagnetic order is identified at $T \le 4.0$ K. 
The exchange interactions are estimated as $J_1$ = 27.8(3)${\mu}$eV and $J_2$ = 1.0(1) ${\mu}$eV, and the saturation field $H_{\rm C}$ is 9.75 T. 
Magnetic excitation measurements with high experimental resolution setup by triple axis neutron spectrometer reveals the instability of one magnon excitation in the field range of $0.7H_{\rm C} \lesssim H \lesssim 0.85H_{\rm C}$. 
\end{abstract}

\pacs{75.10.Jm, 75.25.+z, 75.50.Ee}

\maketitle

Many phenomena in condensed matter science can be explained by using the concept of quasiparticle. 
For example antiferromagnetic order is a result of Bose condensation of magnons~\cite{GiamarchiPRB99,Nikuni00} and superfluidity is those of phonons.\cite{London} 
The quasiparticles in condensed matters, however, can be unstable and decay if allowed by conservation laws. 
Such phenomena was initially predicted in superfluid Helium and was identified by a termination of the excitation at twice the energy of a roton.\cite{Lifsitz} 
The magnon version of the spectral termination was observed in a few quantum magnets, where the one-magnon branch crosses the lower boundary of the two-magnon continuum.\cite{StoneNature,MasudaIPA} 
Recently the instability of magnons with a similar mechanism is predicted in the 2D square lattice Heisenberg antiferromagnets (SLHAF) in high magnetic field.\cite{Zhitomirsky99,Syljuasen08,Syromyatnikov09,Luscher09}
At zero field a two-magnon continuum spreads in the higher energy region for all wave vector ${\bm q}$ and there is no decay channel for one magnon. 
With increasing field the one-magnon branch moves higher around ${\bm q} \sim (\pi ~ \pi )$ and eventually overlaps with the continuum at a threshold field $H^* \sim 0.76 H_{\rm C}$. 
The hybridization of one-magnon with two-magnon continuum induces instability of the one-magnon state. 
While most theoretical calculations including spin-wave,\cite{Zhitomirsky99,Syromyatnikov09} Monte Carlo,~\cite{Syljuasen08} and exact diagonalization~\cite{Luscher09} are for the spin $S=1/2$ case, the field induced magnon instability is universal to 2D and 3D AFMs with general spins. 
It is the change of the curvature of AF magnon dispersion from downward at low field to upward at higher field that qualitatively explains the kinematical instability of the magnon.\cite{Lifsitz,Zhitomirsky99} 

Many model compounds for SLHAF have been reported so far.\cite{Coomer07} 
The required specification for experimental study of the spin dynamics in high field is moderate exchange energy and reasonably low saturation field that can be reached by superconducting magnet. 
From this one may naively infer that a compound with longer inter-magnetic-ion distance, {\it e.g.,} \BaCuGeO\ with 6.0 {\AA}~\cite{Zheludev96} is favorable for experimental investigation. 
Unfortunately, however, the measured $H_{\rm C}$ for \BaCuGeO\ was 27 T~\cite{Nojiri} which is slightly higher than achievable in most superconducting magnets. 
Hence we considered the presumably isostructural Mn$^{2+}$ ($S=5/2$) compound \BaMnGeO\ as a better experimental candidate. 
In this letter we report the experimental observation of a general nature of antiferromagnetic magnon, namely, the field induced magnon instability in $S$=5/2 SLHAF \BaMnGeO\ by combination of bulk susceptibility, high field magnetization, and neutron scattering techniques. 

\begin{figure}
\begin{center}
\includegraphics[width=8.5cm]{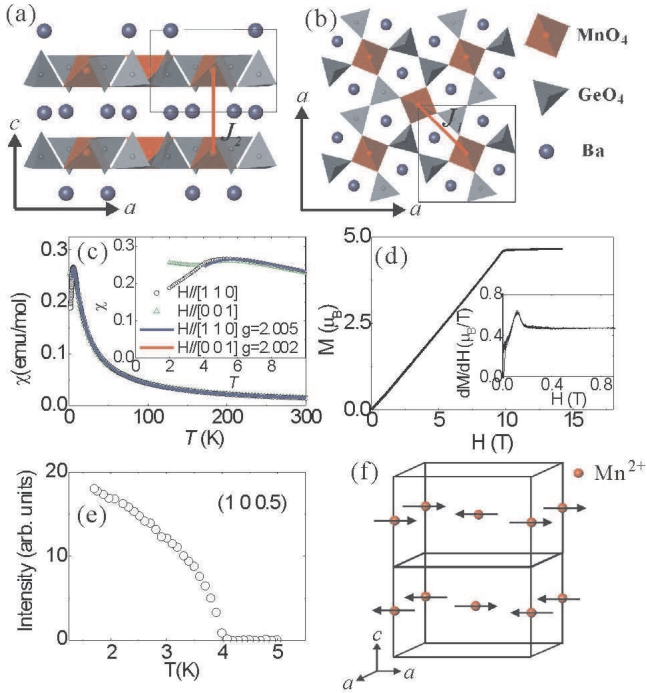}
\end{center}
\caption{(Color online) 
(a) Crystal structure of \BaMnGeO\ projected onto the crystallographic $a$ plane. The solid rectangle indicates one unit cell. (b) Crystal structure projected onto the $c$ plane. (c) Magnetic susceptibility along $H\parallel [1~1~0]$ and $[0~0~1]$. Solid curves are theoretical calculation for $S=5/2$ SLHAFM.\cite{Lines} Both curves cannot be distinguished within the printable resolution. (d) Magnetization curve at $T$ = 0.5 K. (e) Temperature dependence of the magnetic Bragg peak at ${\bm q} = (1~0~0.5)$. (f) Magnetic structure of \BaMnGeO . 
Magnetic unit cell is doubled in the $c$ direction. 
Moments are confined in the $c$ plane. 
} \label{fig1}
\end{figure}

A single crystal with $7 \times 7 \times 80$ mm$^3$ was grown by the floating zone method. 
The crystal structure was determined by using a commercial x-ray diffractometer with area sensitive CCD. This confirmed that the sample has a tetragonal structure with space group $P\bar{4}2_{1}m$ and lattice parameters with $a=8.505$ \AA\ and $c=5.528$ \AA\ which is indeed isostructural to \BaCuGeO .
As shown in Fig.~\ref{fig1}(a) and (b) MnO$_4$ tetrahedra linked by GeO$_{4}$ tetrahedra forms a square lattice in the crystallographic $c$ plane. 
The Mn$^{2+}$-Mn$^{2+}$ distance is 6.014 \AA\ which indicates that the exchange constant may be reasonably small. 
The square lattice is separated by Ba$^{2+}$ ions in the crystallographic $c$ direction and the system is presumed to be 2D-SLHAF. 

Magnetic susceptibilities, ${\chi}$'s, along $[0~0~1]$ and $[ 1~1~0]$ were measured by a commercial SQUID magnetometer as shown in Fig.~\ref{fig1}(c). 
The applied field is 100 Oe. 
At $T \ge 4.0$ K anisotropy is quite small and almost negligible. 
The gyromagnetic ratios of Mn$^{2+}$ ions are obtained $g=2.005$ and 2.002 for $H \parallel [1~1~0]$ and $[0~0~1]$, respectively by room temperature ESR measurement. 
Substantially small anisotropy is observed due to the orbital angular momentum of Mn$^{2+}$ ion for which the high spin state is zero. 
This indicates that spin exchange interactions are purely isotropic in \BaMnGeO .
The broad maximum at $T \sim 6$ K indicates short range AF fluctuation. 
The susceptibility is reasonably fitted by $S=5/2$ 2D-SLHAF~\cite{Lines} with $J$ = 26 ${\mu}$eV (solid curves). 
Anisotropic behavior at $T \le$ 4 K indicates a magnetic ordering. 
$\chi$ in $[1~1~0]$ does not go to zero at $T$ = 0 K. 
While measuring the susceptibility in the $c$ plane, {\it i.e.,} $a^*-a^*$ plane, by rotating the sample around the $c$ axis we observed almost no anisotropy. 
This indicates an easy-plane type AF order. 
High field magnetization along $H \parallel [1~1~0]$ at $T$ = 0.5 K was measured up to 14 T by using the pulsed magnet installed at IMR Tohoku university. 
The magnetization saturates at $H \ge 9.75 {\rm T} \equiv H_{\rm C}$, which is consistent with that estimated from $S=5/2$ SLHAF with $J = 26$ ${\mu}$eV.  
A peak structure ascribed to spin-flop transition  was observed in the field derivative of magnetization at $H \sim 0.12$ T. 
We attribute this to a small anisotropy in the $c$ plane. 

Neutron diffraction measurements were performed at HQR spectrometer in JRR-3M, Japan Atomic Energy Agency (JAEA). 
Magnetic Bragg peaks are observed at $(h~0~(n+1)/2)$ ($n$ is odd) below $T \le 4.0$ K $\equiv T_{\rm N}$ as shown in Fig.~\ref{fig1}(e). 
18 magnetic Bragg reflections were collected in the $a$ plane. 
We found that easy-plane type AF order as shown in Fig.~\ref{fig1} (f) is consistent with the observed Bragg intensities. 
The direction of spins in the $c$ plane is not determined by the unpolarized neutron experiment. 
The magnitude of the moment is estimated ${\mu}_{\rm Mn} = 4.66(6) {\mu}_{\rm B}$. 

\begin{figure}
\begin{center}
\includegraphics[width=9.2cm]{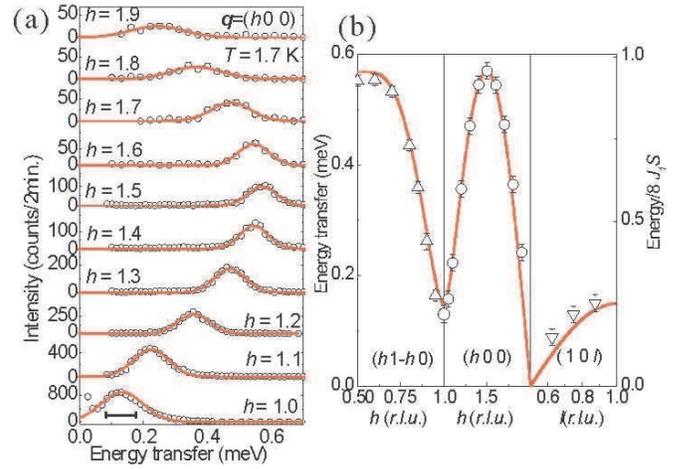}
\end{center}
\caption{ (Color online)
(a) Constant q scans at $(h~0~0)$ at $T$ = 1.7 K and $H=0$T. 
A bar in the bottom shows experimental resolution. 
(b) Dispersion relations in zero field. 
Right axis is scaled by the spin wave band energy $8J_1S$. 
} \label{fig2}
\end{figure}

Inelastic neutron scattering experiments in zero field at $T$ = 1.7 K were performed at HER spectrometer in JRR-3M in JAEA. 
Collimation setup was guide-open-80'-80'. 
A Be filter was installed in front of sample and fixed final energy mode with $E_{f}=3$ meV was used. 
Constant q scans at ${\bm q} = (h~0~0)$ are shown in Fig.~\ref{fig2}(a). 
The bar in the bottom indicates the energy resolution.\cite{Cooper,ZheludevReslib} 
Well defined resolution limited peaks are observed at $\hbar \omega \lesssim 0.6$ meV. 
The excitation energy at ${\bm q} = (1~0~0)$ that corresponds to two dimensional antiferromagnetic wave vector ${\bm q}^{\rm 2D} = ({\pi}~{\pi})$ is 0.1 meV. 
This suggests a weak interplane coupling along the $c$ direction. 
After a series of scans in three directions the dispersion relation is obtained in Fig.~\ref{fig2}(b). 
Enhanced dispersion with the band energy of about 0.55 meV is observed in the $c$ plane. 
Small dispersion in the $c^*$ direction is identified. 
The obtained data is analyzed by classical spin wave theory where the exchange interactions $J_{1}$ and $J_{2}$ in Fig.~\ref{fig1}(a) and (b) are considered in the Heisenberg Hamiltonian, 
\begin{equation}
{\cal H}=J_1\sum _{\langle i,j\rangle }{\bm S}_i \cdot {\bm S}_j + J_2\sum _{\langle k,l\rangle } {\bm S}_k \cdot {\bm S}_{l}.
\label{hamiltonian}
\end{equation}
Here the sums are taken for pairs of spins. 
Obtained parameters are $J_1$ = 27.8(3)${\mu}$eV and $J_2$ = 1.0(1) ${\mu}$eV. 
The value of $J_1$ is consistent with that estimated by susceptibility measurement. 
Thus \BaMnGeO\ is confirmed as a quasi 2D SLHAFM with reasonable energy scale for observing the spin dynamics in high field. 

\begin{figure}
\begin{center}
\includegraphics[width=9cm]{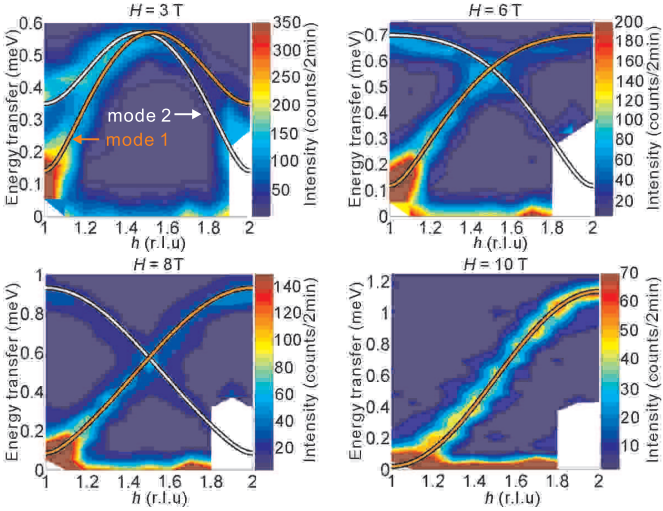}
\end{center}
\caption{(Color online)
False color plots of a series of constant q scans under magnetic fields with wide energy resolution. 
Solid curves are classical spin wave calculation. 
} \label{fig3}
\end{figure}


We performed inelastic neutron scattering with wide resolution for the initial study of high field spin dynamics. 
The data were collected at LTAS spectrometer in JRR-3M in JAEA. 
Collimation setup was guide-80'-open-open. 
A Be filter was installed in front of sample, horizontal focusing analyzer and the $E_{f}=2.6$ meV mode was used. 
The field was applied in the $a$ direction and the temperature of the sample was $T$ = 0.25 K. 
False color plots of constant q scans at ${\bm q} = (h~0~0)$ are shown in Fig.~\ref{fig3}. 
With increasing field the energy of the main spin wave, mode 1, at ${\bm q} = (2~0~0)$ corresponding to ${\bm q}^{\rm 2D} = (0~0)$ increases and the band energy reaches about 1.1 meV at $H = 10$ T ($ \sim H_{\rm C}$). 
The dispersion has a ferromagnetic profile with reduced unit cell in the $a$ plane at $H = H_{\rm C}$. 
In the intermediate field another spin wave, mode 2, is induced due to doubling of the magnetic unit cell in the $c$ direction.  
At $H_{\rm C}$ the doubling is dissolved and the mode 2 disappears. 

Solid curves are dispersions calculated by classical spin wave theory. 
The ground state of canted AF order with tilt angle from easy plane ${\theta} = {\sin}^{-1} (H/H_{\rm C})$ is assumed. 
Dispersion for mode 1 is obtained that 
\begin{equation}
{\hbar}{\omega}_{\bm q}^{1{\rm mag}}=2S \sqrt{(j(0)+j({\bm q}))(j(0)-{\cos}2{\theta}j({\bm q}))}
\label{1magnon}
\end{equation}
where $j({\bm q})$ is Fourier transformation of $J's$. 
The parameters $J_1$ and $J_2$ estimated by zero field data are used. 
The calculation reasonably reproduces the data in all fields. 
In this wide experimental resolution no broadening was observed. 

\begin{figure}
\begin{center}
\includegraphics[width=7cm]{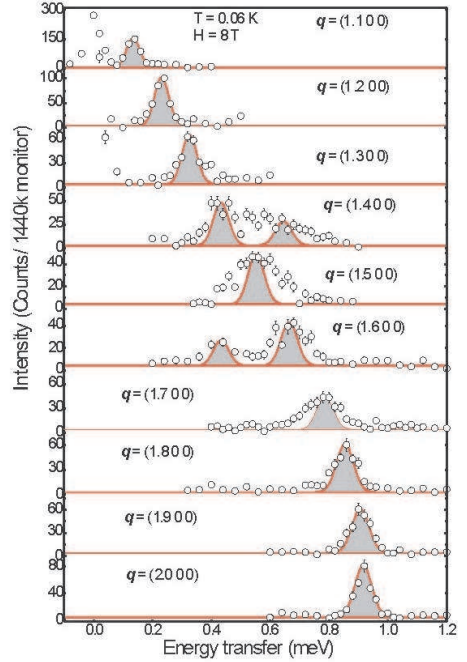}
\end{center}
\caption{(Color online)
Constant q scans at $H$ = 8 T. 
Shaded areas indicates energy resolution. 
} \label{fig4}
\end{figure}

\begin{figure}
\begin{center}
\includegraphics[width=8cm]{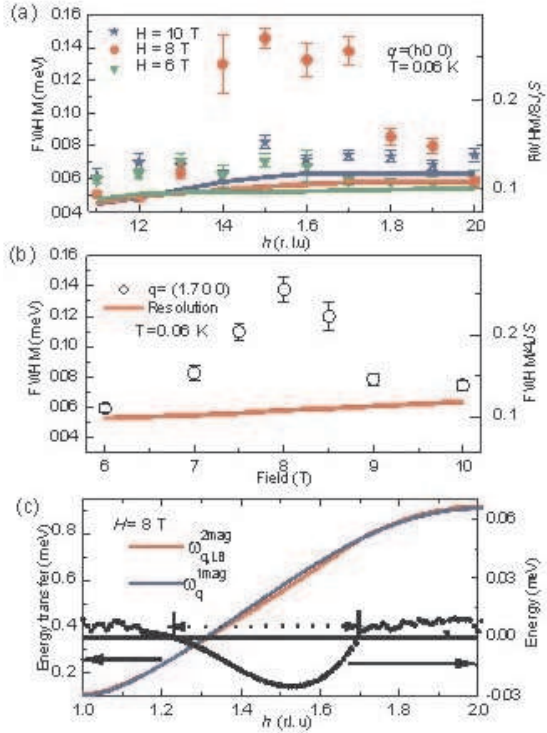}
\end{center}
\caption{(Color online) 
(a) Energy width of the constant q scans at different fields. 
Solid curves indicate the experimental resolution at the peak energies. 
In the right axis the energy is scaled by spin wave band width at zero field $8J_1S$. 
(b) Field dependence of energy width of constant q scans at $q=(1.7~0~0).$ 
(c) Blue curve is one-magnon energy and red curve is the lower boundary of calculated two magnon continuum. Symbols are the subtraction of the latter by the former. 
Dotted arrow indicates the decay channel allowed by conservation laws. 
} \label{fig5}
\end{figure}

To obtain the precise peak profile of the magnetic excitation we performed the measurements with high energy resolution. 
The data were collected at V2/FLEX spectrometer in Helmholtz Zentrum Berlin. 
We put tight collimations, guide-60'-60'-60', with $E_{\rm F}$ fixed at 2.6 meV which gives the best resolution for practical use of 3-axis spectrometer. 
A magnetic field was applied in the $c$ direction. 
Constant q scans at ${\bm q}=(h~0~0)$ with $H$ = 8 T and $T$ = 0.06 K are shown in Fig.~\ref{fig4}. 
At $h \le 1.3$ and $h \ge 1.8$ the excitations are within the calculated experimental resolution~\cite{Cooper,ZheludevReslib} shown by the shaded area. 
Meanwhile at $1.4 \le h \le 1.7$ we clearly observe a broadening of the peaks. 
The shoulder structures at $h$ = 1.4 and 1.6 are a contribution from another mode. 
The energy widths are separately obtained by Gaussian fittings and plotted as function of $h$ for $H$ = 6, 8, and 10 T in Fig.~\ref{fig5}(a). 
Solid curves are the experimental resolution at the peak energies. 
At $H$ = 6 T and 10 T, the widths are resolution limited for all $h$. 
Significant broadening is identified at $1.4 \le h \le 1.7$ at $H$ = 8 T. 
In the field dependence of the width at ${\bm q}=(1.7~0~0)$ in Fig.~\ref{fig5}(b), the broadening is observed at 7 T $\le H \le $ 9 T. 

The field induced instability of the magnon is generally predicted in 2D- and 3D-AFM. 
The necessary condition for the instability is that the one-magnon branch enters into the two-magnon continuum. 
Energy and momentum conservation laws for one-magnon and two-magnon are~\cite{Zhitomirsky99} 
\begin{eqnarray}
{\omega}_{\bm q}^{2{\rm mag}}&=&{\omega}_{\bm q_1}^{1{\rm mag}}+{\omega}_{\bm q_2+Q}^{1{\rm mag}}  \label{2magone} \\
{\bm q}&=&{\bm q_1} +{\bm q_2} + {\bm Q}. 
\label{2magtwo}
\end{eqnarray}
Here ${\bm Q}$ is antiferromagnetic vector. 
We numerically obtained the lower boundary of the 2-magnon continuum, ${\omega}_{{\bm q},LB}^{2{\rm mag}}$, by using Eq.~(\ref{1magnon})-(\ref{2magtwo}) and the obtained parameters $J_1$ and $J_2$. 
At low field the condition ${\omega}_{\bm q}^{2{\rm mag}} > {\omega}_{\bm q}^{1{\rm mag}}$ holds for any ${\bm q}$ and there is no decay channel. 
With increasing field ${\omega}_{\bm q}^{1{\rm mag}}$ goes higher and the decay channel comes to open at $H = H^*$. 
In Fig.~\ref{fig5}(c) at $H$ = 8 T we see that ${\omega}_{{\bm q},LB}^{2-{\rm mag}}$ (red solid curve) is lower than ${\omega}_{\bm q}^{1{\rm mag}}$ (blue solid curve) in intermediate ${\bm q}$. 
For clarity ${\omega}_{{\bm q},LB}^{2{\rm mag}}-{\omega}_{\bm q}^{1{\rm mag}}$ is plotted by filled circles. 
It is identified that a decay channel opens at $1.22 \lesssim h \lesssim 1.70$. 
This $h$ range is totally consistent with $h$ where the broadening of one-magnon is experimentally observed in Fig.~\ref{fig5}(a). 
Hence the observed broadening of the peak profiles is ascribed to the field induced instability of magnon that has been theoretically prediced.\cite{Zhitomirsky99,Syljuasen08,Syromyatnikov09,Luscher09} 

For convenience the energy scaled by spin wave band width in zero field $8J_1S$ is shown in the right axes in Fig.~\ref{fig2}(b), \ref{fig5}(a) and (b). 
The most enhanced broadening is about 25\% of $8J_1S$ at $H=8~{\rm T} \sim 0.8H_{\rm C}$. 
The magnon spectrum of the theoretical calculation for $S$=1/2 case is shown in Fig. 3 of Ref.\onlinecite{Zhitomirsky99}. 
At $H=0.85H_{\rm C}$ a one-magnon centered sharp peak with broad two-magnon continuum is predicted. 
At $H=0.9H_{\rm C}$ one-magnon is mostly decayed in wide ${\bm q}$ and the energy width is about 35\% of $8J_1S$. 
A similar result is also shown in {\it e.g.,} Ref.\onlinecite{Syljuasen08}. 
Note that the energy in the theoretical papers is scaled only by $J$. 
The line width of \BaMnGeO\ is somehow comparable to $S=1/2$ calculations. 
This is rather surprising because $S=5/2$ spin is classical and hybridization of one- and two-magnons is expected to be small. 
Indeed in a short report~\cite{Mourigal} the width for $S=5/2$ SLHAF is calculated $\sim J/10$, only 0.5\% of $8J_1S$ at its best. 
The enhanced instability in \BaMnGeO\ might be related to the crossing of mode 1 and mode 2 that occurs in the same ${\bm q}$ regime in Fig.~\ref{fig3}. 

At $H>H_{\rm C}$ with increasing field the decay channel allowed by conservation law eqs.(\ref{2magone}) and (\ref{2magtwo}) becomes larger in the ${\bm q}$ space. 
Consistently calculations for the $S$ =1/2 case shows that the one magnon decays almost everywhere at $H=0.9H_{\rm C}$.\cite{Zhitomirsky99} 
It is only for $H>0.99H_{\rm C}$ that the sharp one magnon excitation recovers due to the total suppression of quantum fluctuations. 
In \BaMnGeO\ the one magnon recovers much faster as shown in Fig.~\ref{fig5}(b). 
Already at $H=0.9H_{\rm C}$ the peak is within our experimental resolution. 
The restricted field range is presumably because the $S=5/2$ spin is more classical than $S=1/2$. 

To conclude we observed field induced one-magnon instability in $S$=5/2 2D-SLHAF \BaMnGeO\ by inelastic neutron scattering. 
The instability field region is identified as $0.7H_{\rm C} \lesssim H \lesssim 0.85H_{\rm C}$ at ${\bm q}=(1.7~0~0)$. 
The width of the broadened excitation is comparable to theoretically calculated $S=1/2$ case. 
Advanced theoretical study for classical spin would be important. 
Magnon instability is now believed to be ubiquitous in various types of magnet including triangular and kagom\'e lattices.\cite{Chernyshev} 
Further experimental study in new compounds would be interesting. 

This work was partly supported by Yamada Science Foundation, Asahi
glass foundation, and Grant-in-Aid for Scientific Research (No.s
19740215 and 19052004) of Ministry of Education, Culture, Sports,
Science and Technology of Japan.



\end{document}